# Complementary-like Graphene Logic Gates Controlled by Electrostatic Doping

*Song-Lin Li \*, Hisao Miyazaki, Michael V. Lee, Chuan Liu, Akinobu Kanda, and Kazuhito Tsukagoshi \**

[\*]     Dr. S.-L. Li, Dr. H. Miyazaki, Dr. M. V. Lee, Dr. C. Liu, Prof. K. Tsukagoshi
International Center for Materials Nanoarchitectonics (MANA), National Institute for Materials Science, Tsukuba, Ibaraki 305-0044, Japan
CREST, JST, Kawaguchi, Saitama 332-0012, Japan
E-mail: li.songlin@nims.go.jp, tsukagoshi.kazuhito@nims.go.jp
         Prof. A. Kanda
Institute of Physics, University of Tsukuba, Tsukuba, Ibaraki 305-8571, Japan
CREST, JST, Kawaguchi, Saitama 332-0012, Japan

The isolation of graphene offers an emerging candidate to nanoelectronics,[1] because it exhibits a range of remarkable electrical properties, such as high carrier mobility,[2,3] shorter scaling length,[4,5] and compatibility with planar lithography process. Although the ambipolar conduction behavior intrinsic to graphene hinders the feasibility of logic devices directly in a conventional complementary (CMOS) architecture, several methods have been recently developed.[6-11] Among them, a self-adaptive complementary-like architecture based on ambipolar transistors is especially interesting,[12-16] because the ambipolar nature is used as a benefit rather than a drawback to form logic devices. Free of doping, charge neutrality points (CNPs) of two involved transistors are controlled by a supply bias ($V_{DD}$) and the intrinsic $p$- and $n$-conduction branches are delicately combined to construct a complementary geometry. Following this route, we previously demonstrated the first example of graphene voltage inverters with voltage gain greater than one.[10] Enhanced device performance was achieved by introducing a band gap into the channels.[11] It would be interesting for graphene nanoelectronics if component integration and logic devices advanced than inverters can be realized.

With constant miniaturation of chip size, the $V_{DD}$ level in microelectronics has presently scaled down to about 1 V. In an attempt to fabricate advanced logic gates, we found that the self-adaptive architecture has an intrinsic limitation for the integration of complicated structure under low $V_{DD}$ bias,[10-12] because the CNP splitting rely solely on the magnitude of $V_{DD}$ and is rather small as $V_{DD}$ is 1 V or less, which makes the complementary transistor pairs indistinguishable in complicated integration structures with more than two transistors. A direct consequence includes low performance of NAND and NOR gates in which four transistors are involved.[12] Surely, the issues can be solved by impurity dopings on graphene as on conventional silicon, but it will cause severe mobility degradation of the channels since graphene is very sensitive to defects and impurities. In this work, we report a novel method to electrostatically control the CNP positions with modulated gating capacitances and apply it to construct logic gates with semiconducting graphene. The channel regions covered with thick and thin dielectric layers are used as equivalent $p$- and $n$-type components, respectively. This technique allows large CNP splitting at low $V_{DD}$ and continuous tuning of the CNP splitting, which offers great flexibility for component integration. For the first time, complementary-like graphene NOR and NAND gates were demonstrated, which are building pillars for various logic functions. This provides a possible route for future graphene logic circuits.

The fabrication flow and corresponding optical images at each step are shown in figure 1. First, graphene flakes were mechanically exfoliated and transferred onto highly doped silicon substrates with 90 nm thermally grown $SiO_2$ insulating layers (Figure 1a). The numbers of layers of graphene were identified by an optical contrast method and confirmed by Raman spectra (Supporting Information, Figure S1). Pristine large-area graphenes often contain flakes with various numbers of layers (Figure 1b). Thus, standard electron beam lithography and oxygen plasma etching were employed to remove extra graphene flakes and to define appropriate channel patterns (Figure 1c). In order to introduce a band gap[17-20] and increase switching performance of the channels, bilayer graphene (BLG) was adopted. In Figure 1d, three independent devices with six BLG channels in each device were formed for the integration for NOR and NAND gates. After depositing the Au/Ti source and drain electrodes (Figures 1e and 1f), a 1 nm Al and a 4 nm $SiO_2$ were deposited on some graphene channels through resist masks defined by electron beam lithography (Figures 1g and 1h). The ultrathin Al layer is used as both the seed layer for $SiO_2$ nucleation and as a protection layer for residual carrier suppression.[21] The $SiO_2$ layer is used to modulate the top gate (TG) coupling capacitance to form $p$-type components in the complementary-like geometry. Finally, the $AlO_x$ dielectric and Al top gate were simultaneously defined by directly evaporating 30 nm aluminum metal (Figure 1i) with a post-passivation.[22] As shown in Figure 1j, three out of six channels in each device are covered with $AlO_x/SiO_2/AlO_x$ dielectric stacks (Capacitance $C_{TG-p}$), while the remaining three are covered only with $AlO_x$ (Capacitance $C_{TG-n}$). For the TG coupling capacitances, $C_{TG-p}$ is modulated with a smaller value than $C_{TG-n}$ by inserting a $SiO_2$ layer, which is the basis for the electrostatic doping modulation.

It was theoretically predicted that the band gap in biased BLG is given by[18] $\Delta_g = \left[e^2V^2t_\perp^2/(e^2V^2+t_\perp^2)\right]^{1/2}$, where $t_\perp$ and $eV$ represent the interplane hopping energy and the interplane energy difference caused by perpendicular electric fields, respectively. The band structure and resulted band gap for transistors covered with different dielectric layers should be essentially same and has no correlation with top dielectric capacitances. The availability of sizable band gaps and high switching ratios in the pure $AlO_x$ insulated BLG transistors have been demonstrated previously.[23] The same behavior is expected in the $AlO_x/SiO_2/AlO_x$ insulated transistors; we confirmed the presence of band gaps in this configuration, as shown in Figure 2a. The resistance ($R$) peak shows a sharp enhancement with bottom gate voltages ($V_{BG}$) (from 0 to −30 V), consistent with the predicted creation and expansion of the band gap relative to the perpendicular electric field.[17-20] The highest switching ratio is about 100 at $V_{BG}$ = −30 V, which is close to the pure $AlO_x$ insulated transistors. Figure 2b shows the CNP traces for both the $p$- and $n$-transistors to estimate the TG coupling capacitances. According to the relation $C_{TG} = C_{BG} \Delta V_{BG} / \Delta V_{TG}$, the $C_{TG-n}$ and $C_{TG-p}$ are ~920 and ~430 nF cm$^{-2}$, respectively.

The key idea of this work is that the relative positions of CNPs for the equivalent $p$- and $n$-type transistors are electrostatically controlled by different TG coupling capacitances. This can lead to larger CNP splitting and makes the equivalent $p$- and $n$-type transistors more distinguishable on polarities and more effective to mimic the complementary function required by logic devices. The principle for the formation of CNP splitting between transistors is as follows. For a graphene channel initially doped with charge density $n_0$, the gated carriers are correlated with both gates. At CNP, there is a relation between the two gate voltages $C_{TG}V_{TG}^{CNP} = -n_0 - C_{BG}V_{BG}$. Thus, the CNP position $V_{TG}^{CNP}$ can be electrostatically controlled by $V_{BG}$. In the channels with two types of top dielectrics, the charge neutrality condition requires $C_{TG-n}V_{TG-n}^{CNP} = C_{TG-p}V_{TG-p}^{CNP} = -n_0 - C_{BG}V_{BG}$, where $V_{TG-n}^{CNP}$ and $V_{TG-p}^{CNP}$ represent the CNP positions for n- and p-transistors. Under large negative $V_{BG}$, the term $-n_0 - C_{BG}V_{BG}$ is positive, which makes $V_{TG-p}^{CNP} > V_{TG-n}^{CNP}$ due to $C_{TG-p} < C_{TG-n}$ and leads to an electrostatic CNP splitting (horizontal arrows in Figure 2b). Interestingly, both the magnitude and position of the splitting can be tuned by $V_{BG}$, offering an additional control on the resulting complementary region. Together with the original $V_{DD}$ induced splitting, this electrostatic splitting largely increases the total splitting magnitude and makes the complementary transistors more distinguishable in polarity for complicated transistor integration.

The electrostatic doping method is firstly checked on a NOT gate, which is the simplest logic gate, composed of only a complementary transistor pair (Figure 2c). Figure 2d shows typical $R$ curves for each involved transistors. At $V_{BG} = -25$ V and $V_{DD} = 0.6$ V, the two CNPs of the transistors are located around 0.5 V and 1.2 V on the axis of input voltage ($V_{IN}$), giving rise to a CNP splitting of 0.7 V, much larger than the value obtained in the pure $V_{DD}$-driven geometry.[10,11] Figure 2f compares the magnitudes of CNP splitting with and without electrostatic splitting. With electrostatic splitting, the CNP splitting exhibits about 0.6 V enhancement for all $V_{DD}$ at $V_{BG} = -30$ V. Figure 2e presents the voltage transfer curve corresponding to the CNP splitting shown in Figure 2d. Minimum and maximum values for output voltage ($V_{OUT}$) are also located at $V_{IN}\sim 0.5$ V and $\sim 1.2$ V, respectively, which correspond to the CNP positions of the transistor pair. Within the CNP splitting region, a sharp voltage inversion forms due to the variation of resistance ratio between two transistors. Outside the CNP splitting region, a $V_{OUT}$ degradation arises, which is a unique feature of the ambipolar-transistor based complementary-like architecture;[12-16] this feature originates from the overlap of same carrier polarity (pp or nn combination). Two parameters are needed to evaluate the performance of this type of NOT gate. One is the voltage gain which is defined as the maximum value of $-dV_{OUT}/dV_{IN}$, and the other is the output swing defined as $(V_{OUT}^{max} - V_{OUT}^{min})/V_{DD} \times 100\%$.

Here we focused on the logic characteristics at low $V_{DD}$ region ($\leq 1$ V), which is required by device design and integration on nanometer scale. Figure 3a shows the $V_{DD}$ dependence at a fixed $V_{BG}$. As $V_{DD}$ changes from 0.1 to 1 V, the voltage gain increases from 0.4 to 2.5. In Figure 3b with $V_{DD}$ fixed, the output swing is enhanced from 50 to 90% as $V_{BG}$ changes from −15 to −25 V. Figures 3c and 3d systematically summarize the performance dependence on $V_{BG}$ and $V_{DD}$. For most measured conditions, the NOT gate exhibits voltage gain >1 and output swing >60%. The output swing is improved as compared with previous pure $V_{DD}$-driven devices under same biasing conditions.[10,11] Systematically, voltage gain increases with both $V_{BG}$ and $V_{DD}$; whereas output swing increases with $V_{BG}$, but first increases and then decreases with $V_{DD}$. The fact that both parameters increase with $V_{BG}$ is attributed to the enhancement of band gap with $V_{BG}$. The inverse relationship between output swing and $V_{DD}$ in high $V_{DD}$ region is due to the minority carrier injection effect, which increases the off-state current and degrades switching ratio (Supporting Information, Figure S2), as observed in small-band gap CNT transistors.[24,25] Conversely, the effect of minority carrier is trivial in large-band gap CNT transistors with sufficiently large channel/drain Schottky barriers. Minimal effect of minority carrier is anticipated in graphene channels when a large enough band gap is introduced in future.

As mentioned, it is difficult to fabricate high-performance advanced logic gates due to limited CNP splitting and output swing in the pure $V_{DD}$-driven complementary geometry.[12] Fabrication becomes even more problematic with small $V_{DD}$ bias ($\leq 1$ V) where CNP splitting is reduced even more. Here we show that enhanced CNP modulation combined with band gap engineering can improve the output swing performance, up to 90% within the measured bias ranges (Figure 3d). The significant improvement of CNP modulation enables us to construct high-performance dual-input graphene NOR (Figure 4a) and NAND (Figure 4c) gates, which are the first pillar for logic circuits. Panels b and d in figure 4 show detailed logic operations, $V_{OUT} = \overline{A + B}$ for the NOR gate and $V_{OUT} = \overline{A \cdot B}$ for the NAND gate, under different $V_{BG}$ biases. In the notations NOR-XX and NAND-XX, the numbers XX (X = "0" or "1") denote the respective Boolean logic states (*i.e.*, low or high voltage level) for the two inputs. The measured $V_{OUT}$ states for the both logic gates reflect the output Boolean states shown in the truth tables (Figures 4a and 4c). The optimal $V_{BG}$ biases are about −19 and −17 V for the NOR and NAND gates, respectively. According to previous result, the transport band gap is ~100 meV at $V_{BG}$ = −30 V. We estimated the induced transport band gaps are ~60 meV in the channels of the NOR and NAND gates, considering that the band gap is roughly proportional to $V_{BG}$ under moderate electric fields.[26] Despite the relatively small band gap, the signal window reaches as high as 30%–40% of $V_{DD}$.

We note that the inaccessibility of higher $V_{BG}$ and larger band gap here is due to a shift effect of CNPs and complementary region with $V_{BG}$. In present band gap scheme, higher $V_{BG}$ makes larger band gap, but simultaneously over-positive shifts the complementary region, leading to a smaller $V_{OUT}$ response at "0" input state ($V_{IN}$ = 0), as shown in Figure 2e. This situation would be avoided if a $V_{BG}$-independent band gap creation technique is employed, such as structure perforation (graphene nanomesh),[27-29] selective etching,[30,31] or chemical modification.[32,33] In a recent example, using selective etching of the graphene edge, a $10^3$ switching ratio was achieved at room temperature without introducing considerable defects into basal planes.[30] These newly developed techniques, which are lithography compatible, may also lend themselves to high-performance logic devices.

The significance of demonstration of NOR and NAND gates lies in the fact that they are the first pillar for Boolean logic and they can be used to construct all other logic functions in electronic computation (Supporting Information, Figure S3). The presentation of these elementary gates demonstrates the possibility of producing logic circuits on semiconducting graphene and thus represents a landmark towards graphene nanoelectronics. The present work also signifies a simple fabrication technique for logic components. Using the intrinsic *p*- and *n*-conduction branches to form a complementary region described herein does not require external doping by impurities. This helps to preserve the intrinsic high carrier mobility of the channels and to realize high-speed signal transfer. Most importantly, graphene fabrication is lithography compatible, making it superior to device fabrication from nanowires or nanotubes, which require individual alignment on substrates. This work provides a general technique for fabricating advanced logic components and practical circuits from ambipolar materials, including organic semiconductors and carbon nanotube thin films. The band gaps and switching characteristics in such materials are typically better than in pristine graphene and thus better performance in devices fabricated from these materials would also be expected when the same methods are applied.

In conclusion, a facile and general method was developed to construct complementary-like Boolean logic gates from ambipolar conduction materials. It relies solely on electrostatic doping by dielectric coupling layers rather than introduction of chemical impurities. By electrostatically controlling the charge neutrality points, intrinsic *p*- and *n*-conduction branches are delicately combined to construct a complementary geometry. This avoids external doping process and largely simplifies fabrication. For the first time, elementary NOR and NAND logic gates were realized on semiconducting bilayer graphene channels. This method provides a general route in development of logic devices for ambipolar materials and is not limited to graphene.

**Experimental Section**

The samples were mounted on chip carriers and sealed in a vacuum sample tube. During measurement, the sample tube was immersed in liquid nitrogen and the sample temperature was about 79 K. The low temperature environment has two implications. First, it reduces the thermal activation energy and to demonstrate the effect of small band gaps (60-100 meV) in our graphene samples. Second, it allows for applying high electric field on both top and back gates without dielectric breakdown.

**Acknoledgements:** We appreciate Dr. H. Hiura for fruitful discussion. This work was supported in part by a Grant-i*n*-Aid for Scientific Research (No. 21241038) from the Ministry of Education, Culture, Sports, Science and Technology (MEXT) of Japan, and by the Funding Program for World-Leading Innovative R&D on Science and Technology (FIRST Program) from the Japan Society for the Promotion of Science (JSPS).)

**Figures**

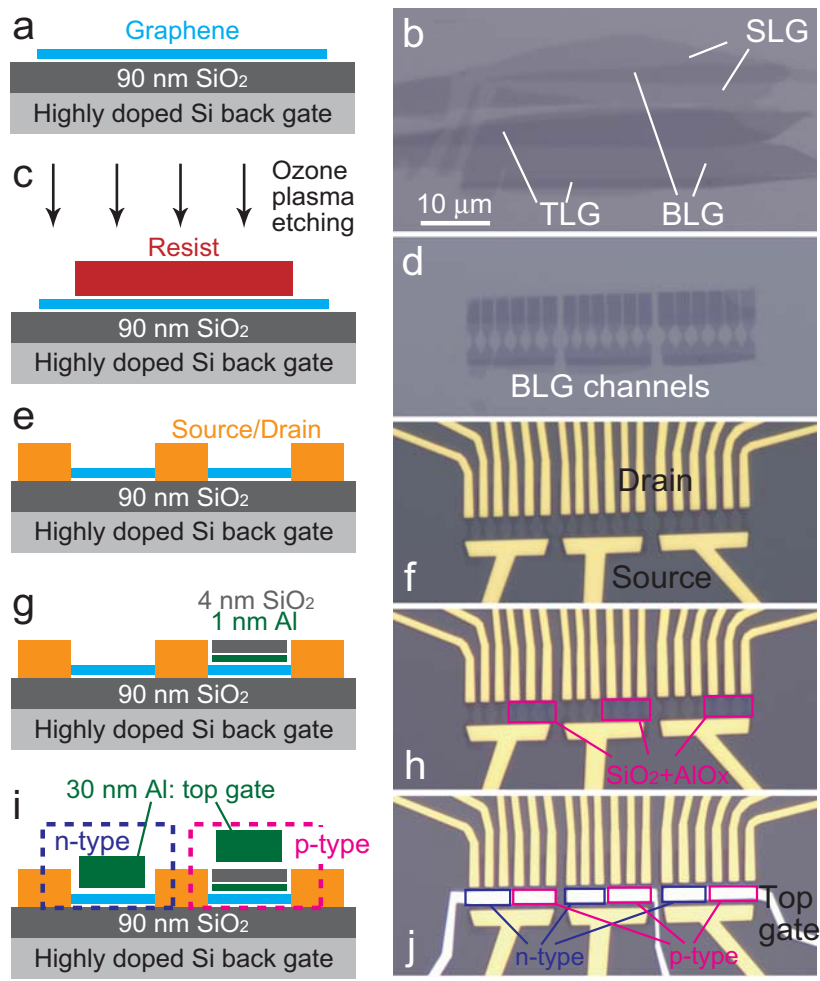

**Figure 1.** Fabrication flow and corresponding optical images at each step. (a) and (b) Exfoliated graphene flakes on 90 nm SiO$_2$ surface. The as-prepared sample contains graphene flakes with different numbers of layers. (c) and (d) Oxygen plasma etching with resist masks defined by electron beam lithography. (e) and (f) Deposition of source and drain electrodes. (g) and (h) Formation of transistors with small top gate coupling capacitance by additionally depositing 1 nm Al and 4 nm SiO$_2$, which serve as the *p*-type component in the complementary-like devices. (i)-(j) Simultaneous formation of Al TG metal and AlO$_x$ dielectric stack by depositing 30 nm metal Al.

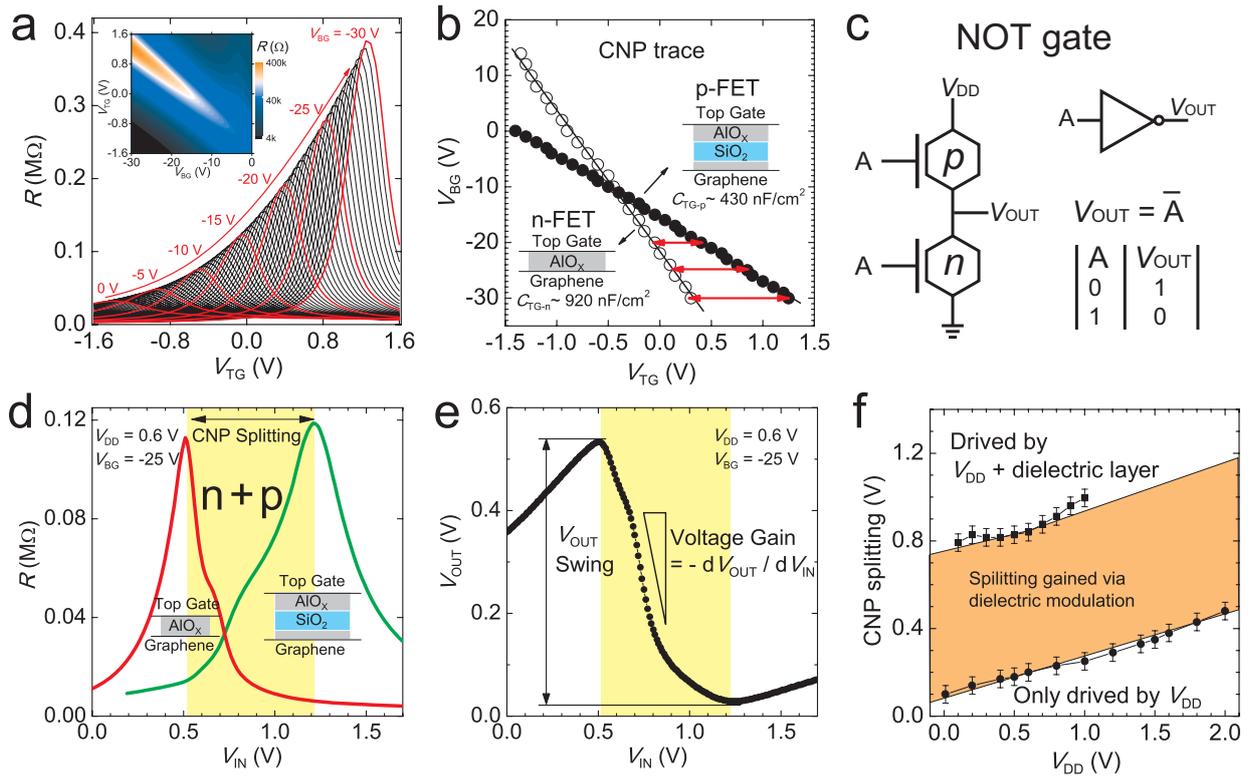

**Figure 2.** (a) *R* curves of a transistor with $AlO_x/SiO_2/AlO_x$ dielectric stacks under different $V_{BG}$ biasing conditions (from 0 to −30 V, 0.5 V step). The dramatic increase of *R* peak is an indication of the expansion of band gap with $V_{BG}$. The inset is the 2D *R* plot of *R* vs. $V_{TG}$ and $V_{BG}$ to show the CNP trace by which the effective $C_{TG}$ can be estimated. (b) Traces of CNP for the BLG transistors insulated by thin (pure $AlO_x$) and thick ($AlO_x/SiO_2/AlO_x$ mixed) dielectric layers. They serve as the equivalent *n*- and *p*-type components in the complementary-like architecture. (c) Schematic diagram, symbol, truth table, (d) *R* curves and (e) voltage transfer characteristic for a graphene NOT gate. (f) Comparison of the magnitude of CNP splitting with and without applying the electrostatic modulation technique. The shaded area indicates the splitting enhancement gained from this technique.

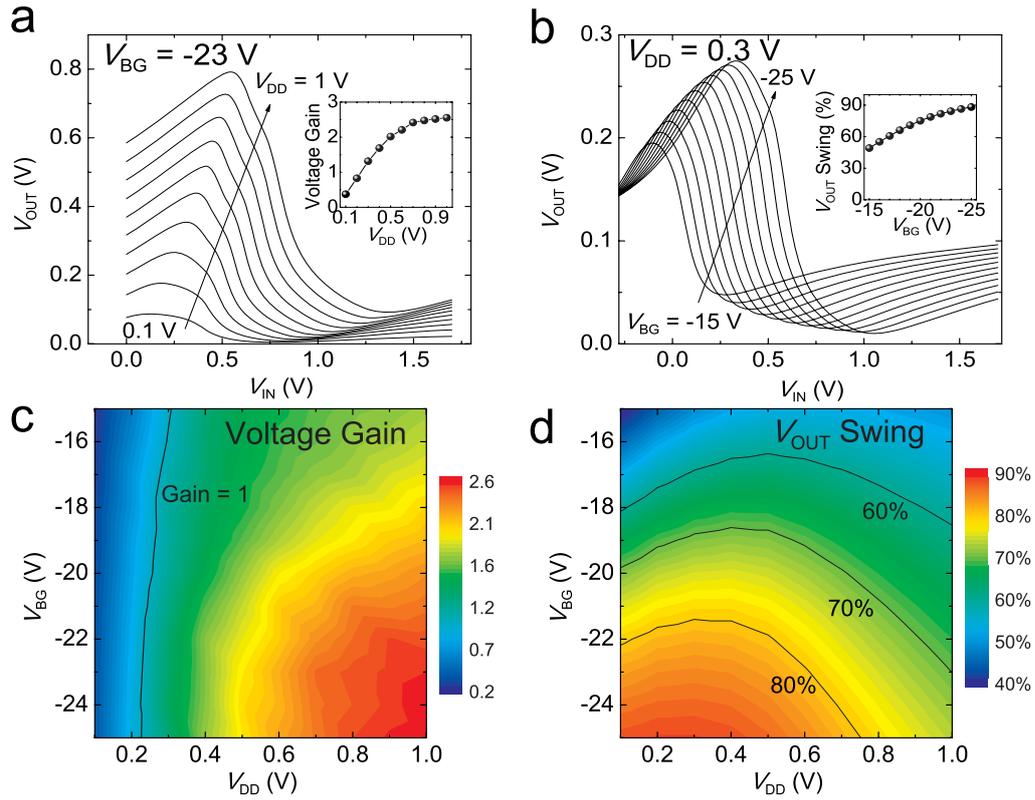

**Figure 3.** Performance of a logic NOT gate. Dependence of the voltage transfer characteristic on (a) $V_{DD}$ (from 0.1 to 1 V) at fixed $V_{BG} = -23$ V and (b) $V_{BG}$ (from −15 to −25 V) at fixed $V_{DD} = 0.3$ V. (c) and (d) The 2D plots of voltage gain and output swing as functions of $V_{BG}$ and $V_{DD}$.

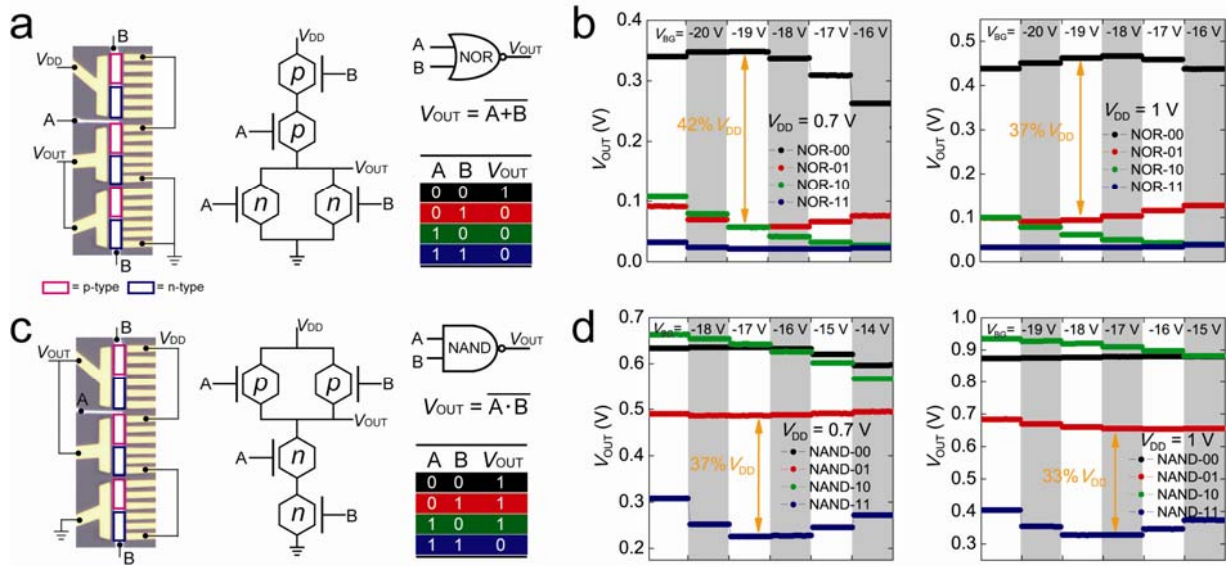

**Figure 4.** (a) Device wiring diagram, schematic diagram, electric symbol, Boolean expression, and truth table for a dual-input NOR gate, in which four transistors including two *p*-type and two *n*-type transistors are used. In the wiring diagram, all the unconnected transistors are floated during measurement. (b) Logic operations for the NOR gate at $V_{DD}$ = 0.7 and 1 V, respectively. The numbers XX (X = "0" or "1") in NOR-XX denote the corresponding logic states of the two inputs. (c) and (d) Corresponding information and results for a graphene NAND gate.

SUPPORTING INFORMATION

# Complementary-like Semiconducting Graphene Logic Gates Controlled by Electrostatic Doping

Song-Lin Li[1], Hisao Miyazaki[1,2], Michael V. Lee[1], Chuan Liu[1], Akinobu Kanda[1,3] and Kazuhito Tsukagoshi[1,2]

[1] MANA, National Institute for Materials Science, Tsukuba, Ibaraki 305-0044, Japan

[2] CREST, Japan Science and Technology Agency, Kawaguchi, Saitama 332-0012, Japan

[3] Institute of Physics, University of Tsukuba, Tsukuba, Ibaraki 305-8571, Japan

**Content**

1. Identification of graphene layer number by optical contrast
2. Origin of bipolar conduction and $V_{DS}$ effect on switching ratio
3. Realization of all logic functions from NAND gates

1. Identification of graphene layer number by optical contrast

According to Ref. R1, the contrast spectra show excellent dependence with layer number up to 10 and exhibit highest resolution for wavelength around 550 nm. After obtaining the optical images under white light (Fig.S1a), we isolated and extracted the corresponding green-channel images, which are used for contrast comparison (Fig.S1b). In Fig.S1b, some areas including intended graphene and suitable substrate nearby were selected (*e.g.*, regions c or d). The substrate region was used as a contrast reference. In the contrast histograms of the selected regions, several peaks appear (Fig.S1 c and d) which correspond to the reflectance contrasts of the $SiO_2$ substrate and graphene flaks. We calculated that the theoretical contrast values for the 1 to 4 layer flakes on 90 nm $SiO_2$ substrates are 94.0, 87.5, 81.1 and 74.6, respectively, normalizing the substrate to 100. In Fig.S1 c and d, the longitudinal axis (green contrast) delivers information about layer numbers, while the horizontal axis (pixel numbers) only reflects the size of the selected areas and thus is irrelated to layer numbers. In Fig.S1c, only an individual graphene peak with contrast ~92 shows up, indicating that the flake in region c is single layer. In contrast, two pronounced peaks near 87 and 81 are shown in Fig.S1d, implying that the region d contains both large 2- and 3-layer flakes.

As an alternative way, Raman spectrum was also employed to confirm the layer numbers determined by the optical method used above. Figure S1e presents the corresponding Raman spectra for the indentified graphene layers. Both the intensity ratios of G band (~1600 cm$^{-1}$) to 2D band (~2700 cm$^{-1}$) and the shape evolution of 2D band show an excellent consistency with the optical method. The optical contrast spectrum is a fast and nondestructive way to identify the layer numbers of graphene flakes.

2. Origin of bipolar conduction and $V_{DS}$ effect on switching ratio

The bipolar conduction behavior in the graphene transistors can be understood within the Schottky barrier transistor model previously proposed for carbon nanotube (CNT) transistors.[R2] The band diagram for such a transistor is schematically shown in Fig. S2c, where the graphene or CNT channel has a band gap $E_g$ ($E_g \geq 0$) and the drain has a Fermi level $E_F$. In transport process, the electrons (e) should overcome a Schottky barrier $\Phi_e$ to reach conductance band ($E_C$) of the channel at the channel/drain interface. Similarly, there is a barrier $\Phi_h$ for holes transport. In the case of large $E_g$, the transistor often becomes unipolar, being p-type when the $E_F$ of drain approaches the $E_V$ of channel ($\Phi_h$~0, $\Phi_e$~$E_g$) and being n-type when $E_F$ approaches $E_C$ ($\Phi_e$~0, $\Phi_h$~$E_g$). The minority carriers are well blocked by the corresponding Schottky barriers and thus well-defined polarity is exhibited.

In the case of graphene channels, the $E_g$ is very small. $E_g = 0$ for single layer (SLG) and $E_g <$ 100 meV for bilayer graphene (BLG). Accordingly, the Schottky barrier is small, because $\Phi_h$, $\Phi_e \leq E_g$. Therefore, the graphene transistors often show bipolar conduction behavior. Only large enough $E_g$ is formed, can a unipolar conduction be obtained. Therefore, it is more accurate to conclude that our graphene transistors operate as normal channel-controlled devices, since the contact barriers cannot efficiently limit carrier injection in small-gap graphene transistors. The conduction characteristics should be understood on the basis of resistance variation of the channels, which is related to the carrier distribution along the channels.

In our logic devices, the performance decreases when $V_{DD}$ is increased. For example, the $V_{OUT}$ swing degrades at high $V_{DD}$ in the inverters (Fig. 3d). Besides, the signal windows of the NOR and NAND gates are also reduced when $V_{DD}$ increases from 0.7 to 1 V (Fig. 3 b-f). Such degradation behavior can be also attributed to the small band gap (< 100 meV) of the channels. To further clarify this behavior, the effect of $V_{DS}$ on an individual FET was measured. The role of $V_{DS}$ for a single FET corresponds to that of $V_{DD}$ for inverters with two FETs, and thus reflects the effect of $V_{DD}$.

Figure S2a shows a series of $I_{DS}$-$V_{TG}$ curves when $V_{DS}$ changes from -0.2 to -2 V. The switching ratio sharply decreased from 50 to 6 as $V_{DS}$ increases, which directly accounts for the degradation of logic performance at high $V_{DD}$. For the transistor, the $I_{DS}$ would scales with $V_{DS}$ if the channel resistance is independent on $V_{DS}$. However, the OFF state current increases by 300 (indicated by the red dots 1 and 2), much higher than the $V_{DS}$ variation scale (-2 V/ -0.2 V =10), suggesting a large $V_{DS}$ effect on the channel resistance. The variation of channel resistance can be more clearly seen in Fig. S2b.

This observed $V_{DS}$ effect can be understood in terms of the charge redistribution model.[R3] When $V_{DS}$ becomes comparable to $V_{TG}$, it produces a substantial position-related potential distribution $V_{Ch}(x)$ along the channel. The effective potential between the TG and channel, $V_{TG,eff}(x) = V_{TG} - V_{Ch}(x)$, is thus strongly dependent on the channel position $x$. At any position, the carrier concentration can be roughly expressed as

$$n(x) = \sqrt{n_0^2(x) + n_g^2(x)} = \sqrt{n_0^2(x) + [(V_{TG,eff}(x) - V_{TG}^0)C_{TG}]^2},$$

where $n_0(x)$ is the density of residual carriers and $n_g(x)$ is the density of gate induced carriers. Fig. S2 c and d schematically depict the carrier distributions for the OFF states shown by the red dots 1 and 2 in Fig. S2a. In the point 2, more carriers are accumulated in the channel due to the higher $V_{DS}$ applied, resulting in a lower channel resistance and a degraded OFF state.

According to the analysis above, we conclude that the adverse $V_{DS}$ effect on logic performance is due to the lack of effective Schottky barriers at drain/channel interface. Larger band gaps and effective Schottky barriers are necessary to achieve high-performance logic devices.

3.  Realization of all logic functions from NAND gates

All other types of Boolean logic gates (*i.e.*, AND, OR, NOT, XOR, XNOR) can be created from a suitable network of NAND gates. Similarly all gates can be created from a network of NOR gates. Historically, NAND gates were easier to construct from silicon MOS technology and thus NAND gates served as the first pillar of Boolean logic in electronic computation.

**Figures**

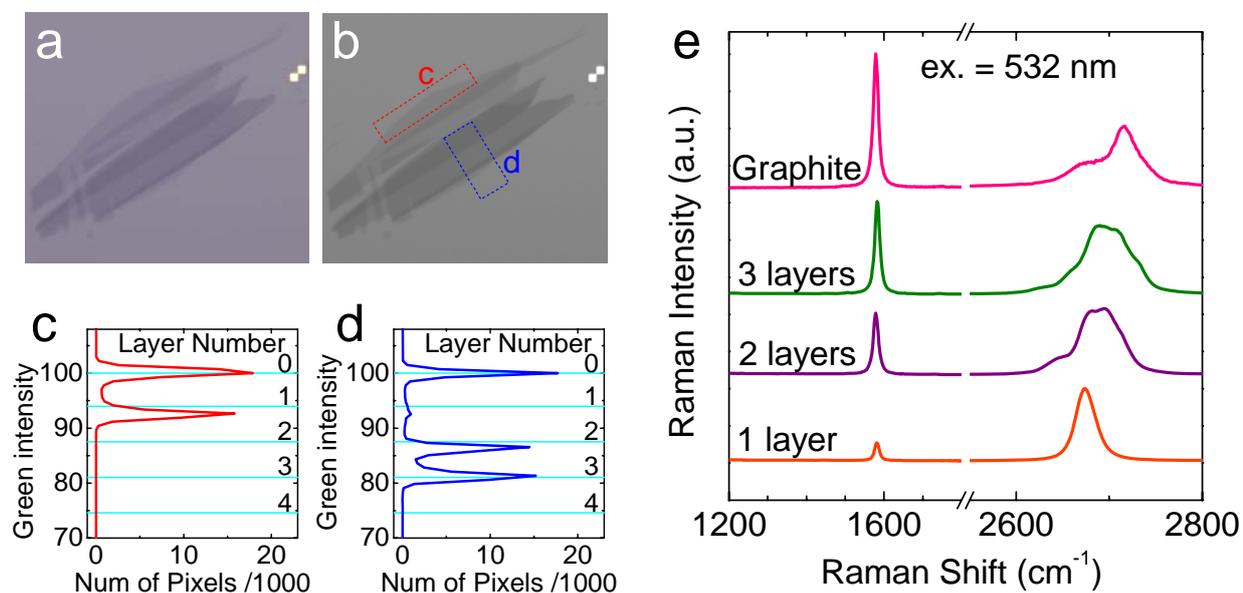

**Figure S1.** Optical images with (a) all RGB channels and (b) individual green channel for graphene samples with multiple flakes on 90 nm $SiO_2$ substrates. The square gold markers are 2 μm in dimension. (c) and (d) show the histograms of green contrast for the regions c and d selected in (b). The substrate contrast is normalized to 100. (e) Raman spectra for graphene flakes with 1-3 layers and a graphite flake.

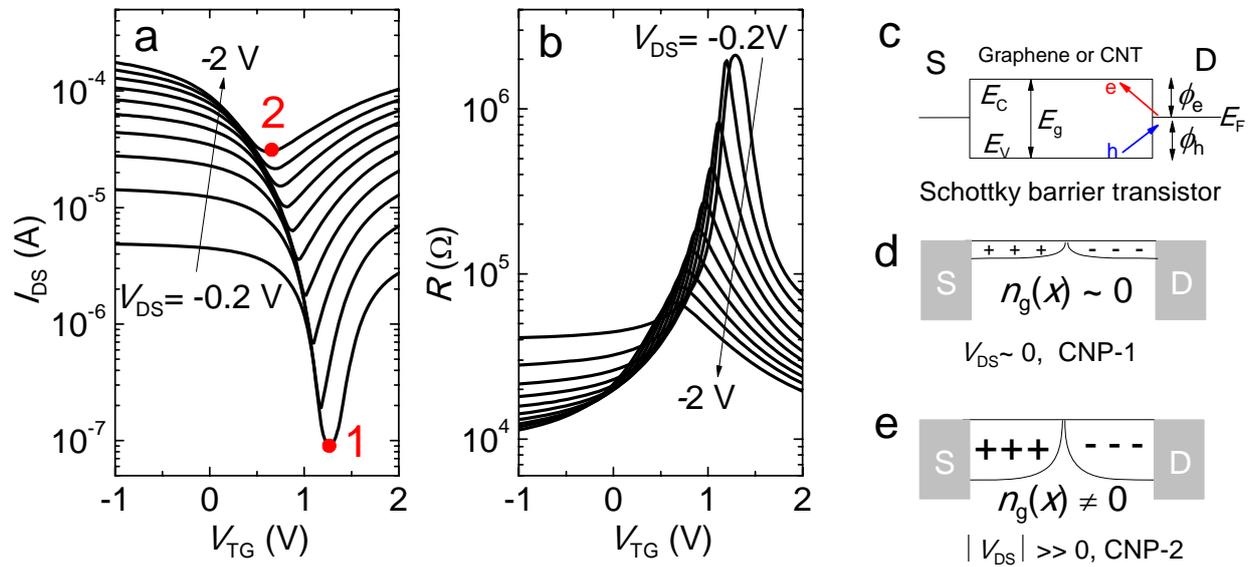

**Figure S2.** (a) Channel current and (b) resistance versus $V_{TG}$ at different drain voltage $V_{DS}$ from -0.2 to -2 V (Step = -0.2 V). (c) Schematic diagram of band lineup for a graphene or nanotube transistor. (d) and (e) Schematic carrier distributions in graphene channels with different $V_{DS}$ biases. The higher the absolute values of $V_{DS}$, the more carriers are induced and the lower channel resistance is obtained.

| Gate | Symbol | Diagram | Number of NAND employed |
|---|---|---|---|
| NOT | 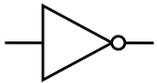 | 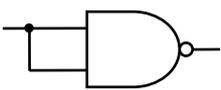 | 1 |
| AND | 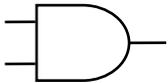 | 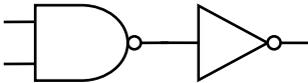 | 2 |
| OR | 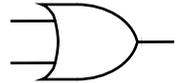 | 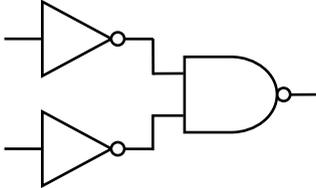 | 3 |
| XOR | 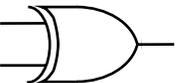 | 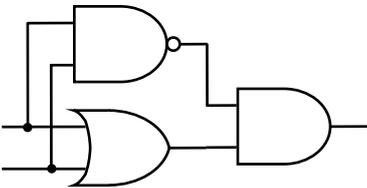 | 6 |
| XNOR | 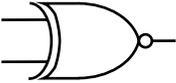 | 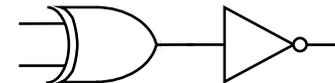 | 7 |

**Figure S3.** The electrical diagram for combination of other types of Boolean logic gates (*i.e.*, AND, OR, NOT, XOR, XNOR) from NAND networks.